\documentclass{mem}
\pdfoutput=1
\usepackage{natbib}
\usepackage{txfonts}
\usepackage{balance}
\usepackage{graphicx}
\usepackage[a4paper]{hyperref}
\idline{75}{282}
\begin{document}
\def\kms{$\mathrm {km s}^{-1}$}

\title{
Double-Barred Galaxies
}

   \subtitle{}

\author{
Peter \,Erwin\inst{1,2} 
}
\offprints{P. Erwin}

\institute{
Max-Planck-Institut f\"{u}r extraterrestrische Physik,
Giessenbachstrasse,
D-85748 Garching, Germany
\and
Universit\"{a}ts-Sternwarte M\"{u}nchen,
Scheinerstrasse 1,
D-81679 M\"{u}nchen, Germany
\email{erwin@mpe.mpg.de}
}

\authorrunning{Erwin }

\titlerunning{Double-Barred Galaxies}

\abstract{
I present a brief review of what is known about double-barred galaxies,
where a small (``inner'') bar is nested inside a
larger (``outer'') bar; the review is focused primarily on their
demographics and photometric properties.  Roughly 20\% of S0--Sb
galaxies are double-barred; they may be rarer in later Hubble types.
 Inner bars are typically $\sim$500 pc in radius ($\sim$12\% the size of
outer bars), but sizes range from $\sim$100 pc to $> 1$ kpc.  The
structure of at least some inner bars appears very similar to that of
outer bars (and single large-scale bars).  Direct and indirect evidence
all support the hypothesis that inner bars rotate independently of outer
bars, although actual pattern speeds for inner bars are poorly
constrained.  Finally, I note that inner bars do not appear to promote
nuclear activity.
}
\maketitle{}

% Section 1
\section{Introduction}

Evidence for double-barred galaxies -- galaxies where a smaller,
\textit{secondary} bar is nested inside a larger, \textit{primary} bar
-- dates back to the mid-1970s, when \citet{dev75} pointed out three
examples (NGC 1291, 1326, and 1543). These and a handful of other
candidates identified in the 1980s \citep[e.g.,][]{kormendy82} remained
isolated observational curiosities for some years.

Theoretical interest was sparked by \citet{shlosman89}, who suggested
that a system of nested bars could be an especially effective way to
transfer gas from galactic (kpc) scales down to near-nuclear (sub-100
pc) scales and thereby feed active nuclei.  They also sketched a
possible formation scenario, in which the gas inflow driven by a
large-scale bar gave rise to a central concentration which became
dynamically decoupled and unstable, leading in turn to the formation of
an independently rotating inner bar.

The early 1990s saw the first attempts to define samples of double bars
and to investigate some of their properties, particularly in work by
\citet{bc93} and \citet{friedli93}; this probably marks the point at
which double bars became recognized as a distinct category of galaxy.
This led in turn to imaging surveys specifically aimed at
identifying and characterizing double-barred galaxies, starting with
\citet{wozniak95} and \citet{friedli96b}, and then with larger samples
by \citet{jungwiert97}, \citet{erwin02}, and \citet{laine02}. Kinematic
studies of double-barred systems are also becoming more common
\citep[e.g.,][]{emsellem01,schinnerer01,moiseev04}; see the contribution
by Alexei Moiseev in this volume.

The same period saw the first detailed attempts to characterize and
model double-barred galaxies -- and their formation -- theoretically
\citep{pfenniger90,friedli93,combes94}.  More recent work has focused on
the question of whether and under what circumstances self-consistent
orbital structures supporting two bars can exist
\citep[e.g.,][]{witold97,witold00,el-zant03,witold08}, and on
hydrodynamical and $N$-body modeling of double-bar formation
\citep[e.g.,][]{rautiainen99,englmaier04,debattista07}. (See the
contributions by Witold Maciejewski and Juntai Shen in this volume for
more details.)

In this review, I will focus on the current observational status of
double-barred galaxies, with a particular emphasis on what know about
them as a population, and what we can tell about the \textit{inner}
bars, both structurally and dynamically.  I will also consider the
question, first raised by \citet{shlosman89}, of whether double-barred
galaxies can promote nuclear activity.

%Kinematic studies: Emsellem+2001; Schinnerer+2001; Moiseev+2002; Corsini+2003; etc.
%
%Theoretical interest:
%
%(Pfenniger \& Normen 1990)
%First numerical models of formation [Friedli \& Martinet; Combes 1996],
%which suggested they might be rather short lived.
%
%Orbital analyses (Pfenniger \& Norman 1990; Maciejewski \& Sparke 1997,
%2000; El-Zant \& Shlosman 2004; Maciejewski \& Athanassoula 2007, 2008)
%
%N-body simulations without gas (e.g., Rautianen \& Salo 1999; Debattista \& Shen 2007); simulations with gas (Englmaier \& Shlosman 2004; Heller+2007, OTHER RECENT SHLOSMAN STUFF?)

% ? FIGURE: one or two samples of DB galaxies, e.g. NGC 2950
%
\begin{figure*}[t!]
%\resizebox{\hsize}{!}{\includegraphics[clip=true]{n2950_example.eps}}
\resizebox{\hsize}{!}{\includegraphics[clip=true]{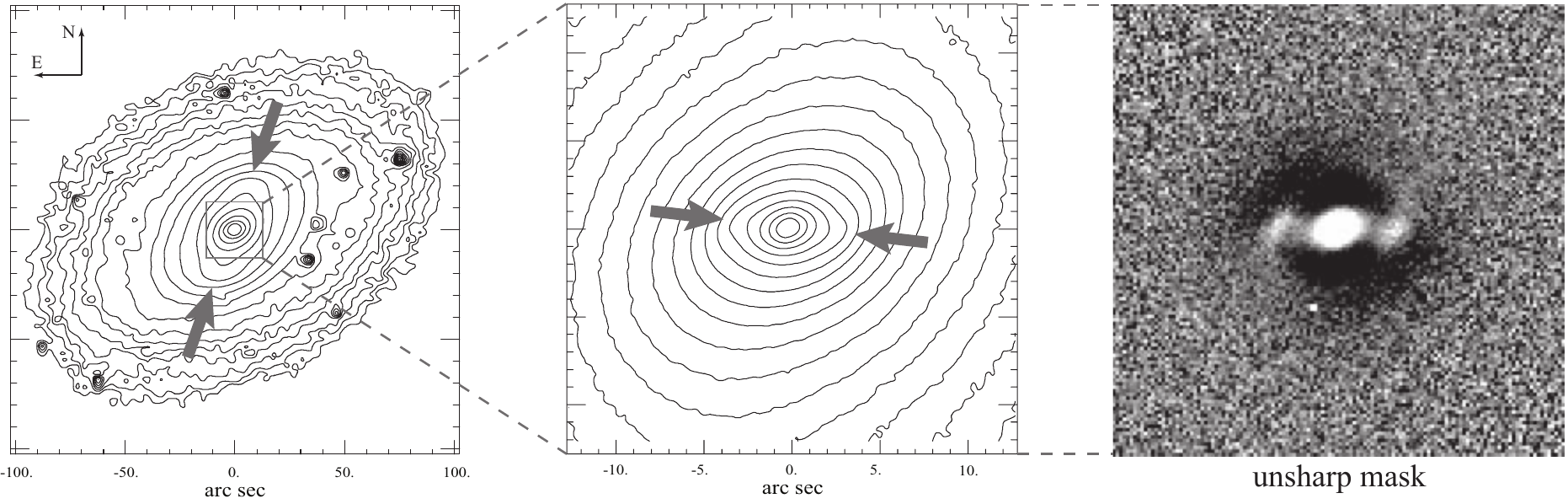}}
\caption{\footnotesize
NGC 2950, an example of a double-barred galaxy.  Left and middle panels show $R$-band isophotes, with the outer bar (left panel) and inner bar (middle panel) indicated by arrows. Right panel: unsharp mask of the same image, highlighting the ends of the inner bar. Although the two bars are (almost) perpendicular in this galaxy, inner and outer bars are found at all possible 
relative orientations (see Figure~\ref{fig:pa}).}
\label{fig:n2950}
\end{figure*}
%

% *Nearest* DB galaxy is NGC 4736 [D=5.2 Mpc, Tonry+2001], or possibly M81
% [Gutierrez et al. 2009; Erwin \& Debattista 2009; D = 3.6 Mpc from 
% Freedman+2001]
% ... Several suggestions that the Milky Way
% itself, now generally accepted as having a large-scale bar, may also
% have a separate inner bar and this be double-barred [REFs, including
% Padova contribution by N.J. Rodriguez-Fernandez]
% 
% Detected out to z = 0.15? [Lisker+2005]; latter suggest that some DBs could
% be detectable out to z ~ 0.5
%
% -- or does this/these go in demographics?

% Section 2
\section{Demographics}

In this section, I consider what we can tell about the general population
of double-barred galaxies and the bars within them: how common they are,
what type of galaxies they are, how large (or small) the bars are, etc.
This is based primarily on two datasets.  The first is an expanded
version of the \citet{erwin-sparke03} sample, which now
encompasses barred and unbarred galaxies of Hubble types S0--Sb with
redshifts $< 2000$ \kms{} and major-axis diameters $> 2.0\arcmin$.  The
other source is an updated version of the database published in
\citet{erwin04}, which attempts to keep track of all well-defined
double-barred galaxies in the literature.  The latter dataset now
includes 61 galaxies; it has the disadvantage of being drawn mostly from
an extremely heterogeneous set of observations, and so is subject to a
variety of poorly known selection effects.

\begin{figure*}[t!]
%\resizebox{\hsize}{!}{\includegraphics[clip=true]{db_fractions_fig.eps}}
\resizebox{\hsize}{!}{\includegraphics[clip=true]{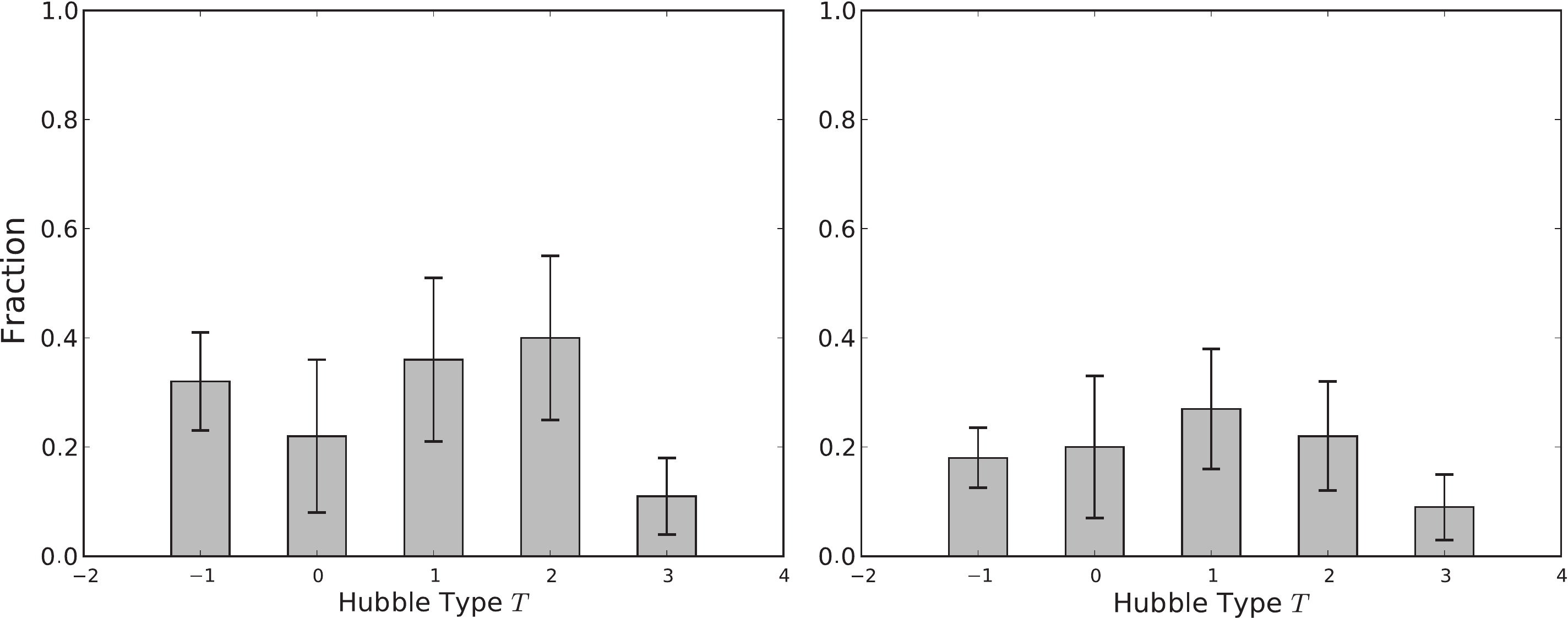}}
\caption{\footnotesize
Fraction of barred galaxies (left) and all galaxies (right) which are
double-barred, as a function of Hubble type.}
\label{fig:fractions}
\end{figure*}

In Figure~\ref{fig:fractions} I show the double-bar fraction as a
function of Hubble types, based on the aforementioned local
sample (\nocite{erwin-sparke03}Erwin \& Sparke 2003; Erwin et al., in
prep). The double-bar frequency is roughly constant at $\sim$30\% of
barred galaxies, or $\sim$20\% of all galaxies, from S0 down through
Sab.  There is \textit{some} evidence that the fraction is smaller for
Sb galaxies.

Are double bars basically a phenomenon of early-type disks?
Unfortunately, we lack systematic surveys of later Hubble types; such
surveys would need to be in the near-IR to avoid dust extinction, and
would need to have high spatial resolutions, since bars in late-type
spirals are systematically smaller than early-type bars \citep{erwin05}.
The current version of the \citet{erwin04} catalog does have several
double-barred Sbc galaxies, but only \textit{one} confirmed double bar
with a late Hubble type (NGC 6946, Sc). While this suggests a lower
frequency of double bars in late Hubble types, the imaging surveys which
provided most of the double-bar detections to date have historically
been heavily biased towards Hubble types earlier than Sc, in part
because many of these surveys have been aimed at Seyfert galaxies (and
matched non-active galaxies), which are primarily early types.

\begin{figure*}[t!]
%\resizebox{\hsize}{!}{\includegraphics[clip=true]{inner-bar-size-histograms.eps}}
\resizebox{\hsize}{!}{\includegraphics[clip=true]{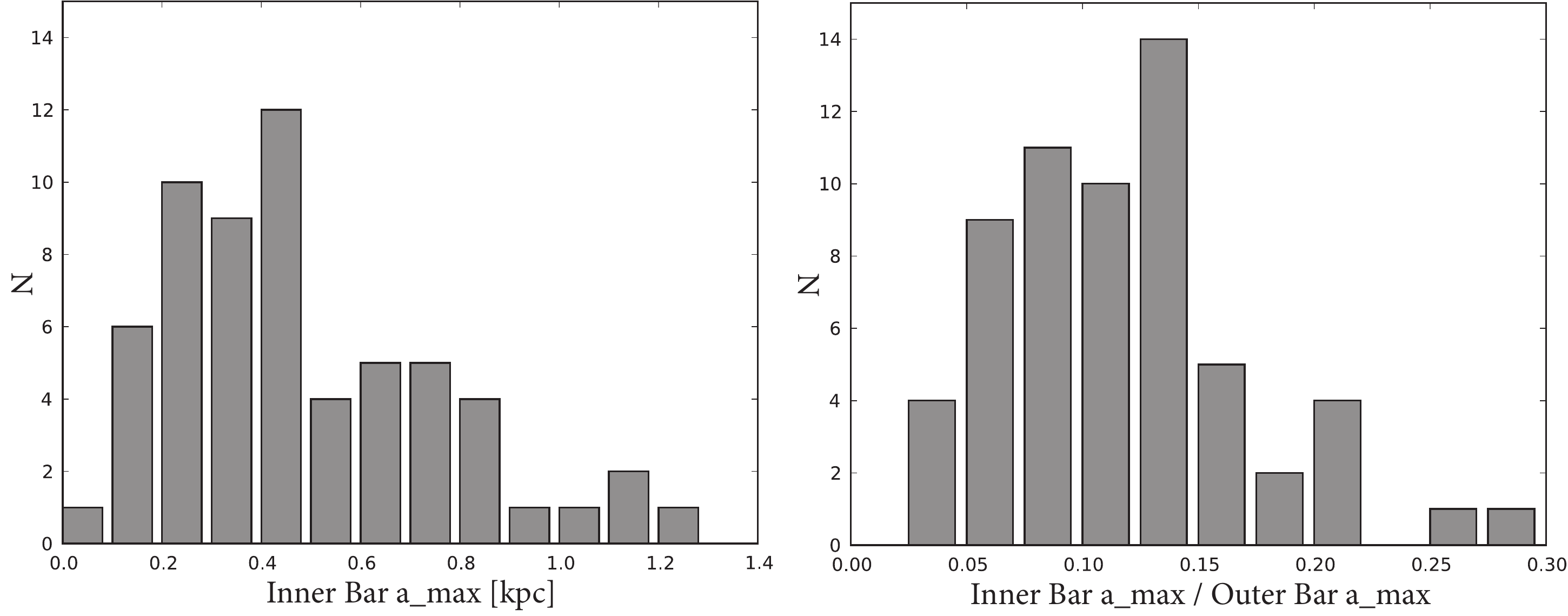}}
\caption{\footnotesize
Left: Absolute semi-major axis sizes of inner bars.  Right: Relative semi-major axis sizes (inner bar size as a fraction of outer bar size).  All sizes are
deprojected.}
\label{fig:sizes}
\end{figure*}

Figure~\ref{fig:sizes} shows the absolute and relative sizes of
\textit{inner} bars from the current catalog of double-barred galaxies. 
In absolute size, inner bars span about an order of magnitude, with
semi-major axes ranging from $\sim$100 pc all the way up to 1.2 kpc; the
median size is $\sim$500 pc (relative to $R_{25}$, the range is
0.01--0.10, with a median of 0.04).  The sizes of inner bars, while
almost always smaller than outer or single bars in early-type disks,
thus actually overlap with the low end of single-bar sizes in
\textit{late}-type spirals (Sc and later; see \nocite{erwin05}Erwin
2005).

Inner bar size \textit{does} correlate with outer bar size, though not
very strongly (Spearman $r = 0.57$).  The median size ratio is $\approx
0.12$; the true median may be slightly smaller, since resolution limits
mean that small inner bars are less likely to be identified. There is a
fairly clear upper limit of $\sim$0.25, which is at least roughly
consistent with theoretical arguments that inner bars cannot be too
large without disrupting the orbits which support the \textit{outer} bars.

The local S0--Sb sample mentioned above contains 55 single-barred and
21 double-barred galaxies. Are there systematic differences between the
two types, which might help use understand why some galaxies have two
bars and other have just one? The answer, for the most part, is no:
single- and double-barred galaxies appear to be very similar in their
global properties (e.g., absolute magnitude, rotation velocity, central
velocity dispersion). The only genuinely significant difference appears
to be in the sizes of the large-scale bars: the outer bars of
double-barred systems are \textit{longer} (typically $\sim$4 kpc in
radius) than the bars of single-barred galaxies (typically $\sim$2.5
kpc).

%

% ? FIGURE: Relative sizes of inner and outer bars vs Hubble sequence

\section{The Structure of Inner Bars}

Are inner bars simply miniature versions of large-scale bars, or are
they a different type of beast altogether?  Theoretical arguments and
models suggest that inner bars should rotate relatively slowly, with
corotation well outside the end of the bar. Thus, they should differ
dynamically from typical large-scale bars in early-type disks (where
most double bars are found); these large-scale bars tend to be ``fast''
in the relative sense, with corotation at or just beyond the end of the
bar.  So we might wonder if this hypothesized difference in relative
speeds is reflected in a difference in structure. In addition, as more
realistic models of inner bars emerge (e.g., from $N$-body simulations),
there is some hope that we can test these models by comparing their
stellar structure with that of real inner bars.

What is curious, then, about the gross photometric structure of inner
bars is that it is often rather similar to that of large-scale bars.
A crude comparison can be had via the use of unsharp masks
\citep[e.g.,][]{erwin-sparke03,erwin04}, which suggests that inner bars
have distinct ``ends'' (i.e., regions where the surface brightness
steepens abruptly), and that some inner bars may have rather high axis
ratios.

\begin{figure*}[t!]
%\resizebox{\hsize}{!}{\includegraphics[clip=true]{n2950-profile-fig.eps}}
\resizebox{\hsize}{!}{\includegraphics[clip=true]{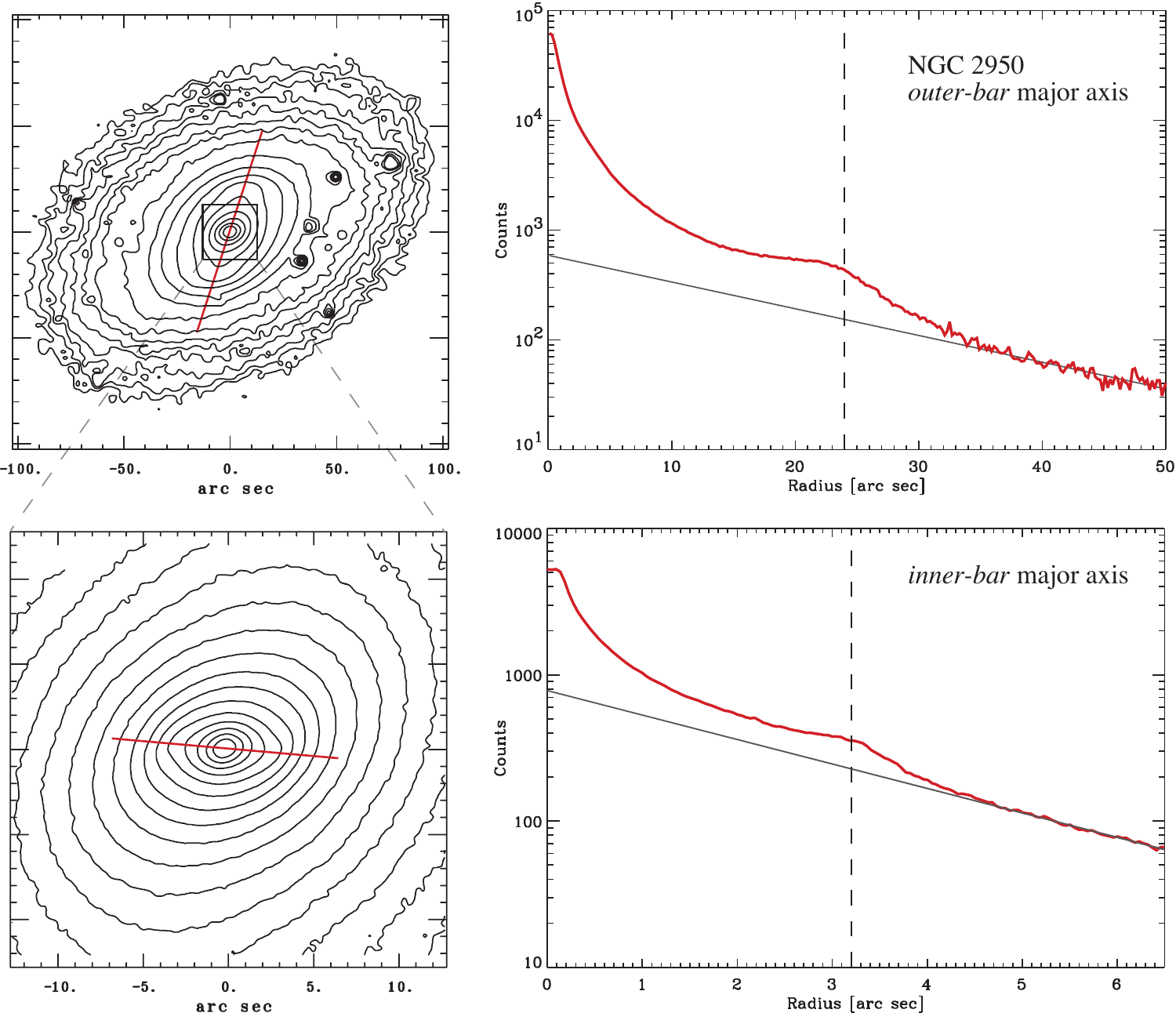}}
\caption{\footnotesize
Profiles along major axes of outer and inner bars of NGC 2950.  Because the bars are (almost) perpendicular, it is easier to trace their profiles independently.  Both bars have remarkably similar profiles --- excellent examples of so-called ``flat'' profiles first identified for large-scale bars by \citet{elmegreen85}.  Vertical dashed lines mark the radius of maximum isophotal ellipticity for each bar.}
\label{fig:n2950-profiles}
\end{figure*}

This can be seem more directly by comparing
surface-brightness profiles along the bar major axis.
Figure~\ref{fig:n2950-profiles} does this for NGC 2950, where the
profile of the inner bar appears to be a scaled-down replica of the
outer bar: a relatively shallow profile (extending out of the
bulge-dominated region) which breaks at a certain radius and then falls
off more steeply, gradually blending into the disk outside.  These are
classical examples of what \citet{elmegreen85} first identified (for
large-scale bars!) as ``flat'' bar profiles, which are common in
early-type disks but rare in late types, where the bar profile is
usually a steep exponential.

% FIGURE: major-axis profiles of inner and outer bars of NGC 2950

%Fourier analysis  
%
%% ? FIGURE: Fourier profiles for NGC 1543 and/or 2859
%
%% ?? FIGURE: A4/A2 profiles??  -- may be better just to mention in text
%
%There is only so much information about inner-bar structure that can be gleaned form major-axis profiles, or from azimuthal profiles which are strongly contaminated by bulge and/or inner disk light.  What would be better would be to isolate the inner bars in some fashion, so that we can study their two-dimensional structure.  This is possible for large-scale bars \citep[e.g.,][]{ohta90}, but more difficult for inner bars, embedded as they are within central disks \citep[e.g.,][]{erwin03-id} and bulges.  Nonetheless, it may be possible to make progress in this direction; some preliminary steps are outlined below.

\subsection{Dissections}

Just as Padova (the site of this workshop) was where Galileo undertook
much of his pioneering research --- which helped break the dogma of
Classical (Aristotelean and Ptolemaic) physics --- it was also the place
where the anatomist Andreas Vesalius helped break the dogma of Classical
medicine, by performing a ground-breaking series of dissections,
culminating in \textit{De humani corporis fabrica} \citep{vesalius}. 
Taking this as inspiration, then, I present selections from an ongoing
project aimed at ``dissecting'' several double-barred galaxies.

\begin{figure*}[t!]
%\resizebox{\hsize}{!}{\includegraphics[clip=true]{dissection-fig.eps}}
\resizebox{\hsize}{!}{\includegraphics[clip=true]{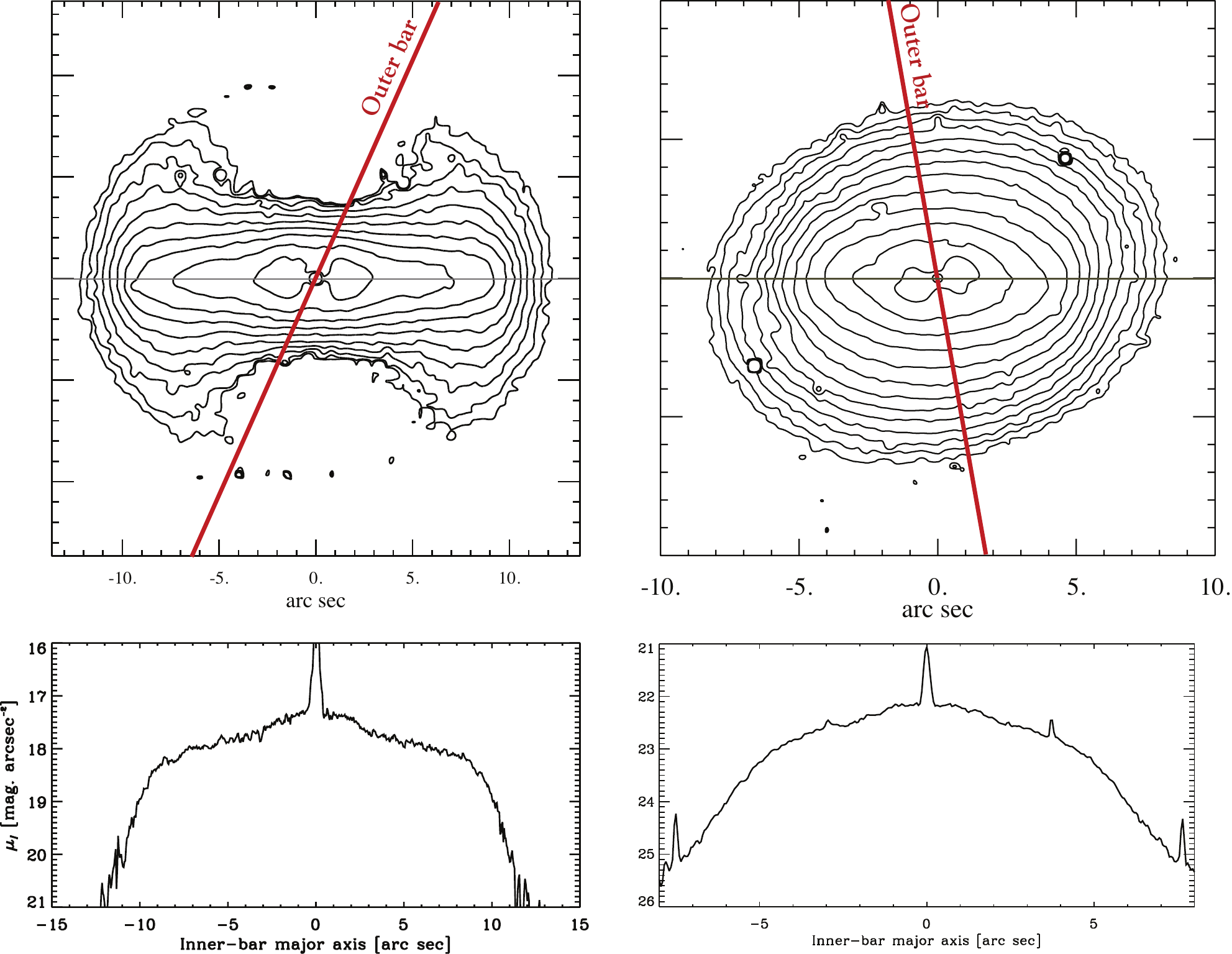}}
\caption{\footnotesize
Isolated inner bars of NGC 1543 (left, WFPC2 F814W) and NGC 2859 (right, ACS/WFC F814W).  All other galaxy components except stellar nuclei have been subtracted. Upper panels show logarithmically scaled isophotes; red lines indicate outer-bar position angle for each galaxy. Lower panels show cuts along the major axis of each inner bar; both profiles are ``flat'' in the sense of \citet{elmegreen85}.}
\label{fig:n1543}
\end{figure*}

The main approach is modeled on that of \citet{ohta90}, who extracted
large-scale bars from several early-type spirals by modeling and
subtracting the disk and bulge, with the residue being the bar proper. 
An elaboration of this process applied to NGC~1543 and NGC~2859 (Erwin
2009, in prep) produces two isolated inner bars
(Figure~\ref{fig:n1543}). That of NGC~1543 is quite narrow (axis ratio
$\approx$ 4:1), and strikingly similar to at least some large-scale bars
(e.g., several of those in Ohta et al.); evidence for a very faint
stellar nuclear ring can be seen.  In NGC 2859, the bar is less
elongated and is embedded in a region of twisted isophotes outside,
possible forming a lens.  As the profiles make clear, both bars are
indeed ``flat'' in the sense of \citet{elmegreen85}.  Although one
should not generalize too freely from a sample of two, it is curious
that at least some inner bars do resemble scaled-down outer (or single)
bars quite closely, given that the theoretical expectation is that they
should differ dynamically.

% FIGURE: NGC 1543 isolated inner bar + profile
%

As a bonus, we can use the extracted bars as direct estimates of how massive these inner bars are --- i.e., what fraction of the total stellar light (and thus stellar mass) they make up.  This turns out to be $\sim$4\% in the case of NGC 1543 and $\sim$10\% for NGC 2859; put another way, the inner bar is $\sim$7\% of the outer bar's mass in NGC 1543, but it is $\sim$25\% of the outer bar's mass in NGC 2859.  These latter values can potentially test models of double-barred galaxies, since theoretical arguments suggest that an inner bar must be massive enough to to produce orbits which can support it, but cannot be \textit{too} massive or it will disrupt the orbits making up the \textit{outer} bar \citep[e.g.,][]{witold00,el-zant03,witold08}.

%POSSIBLE TEXT: EPHEMERAL VS LONG-LIVED:\\
%The fact that double bars are as common in S0 galaxies as they are in spirals already hints that they are not associated with gas or recent star formation.  Multicolor (and near-IR) observations indicate that inner bars are typically \textit{not} made up of young stars [Friedli+96? Erwin+Sparke02?], but are rather composed of old stars like those of the inner disk.  The absence of dust lanes and gas [Erwin+Sparke02; Petitas \& Wilson] in some double-barred S0 galaxies indicates that \textit{if} significant gas inflow is necessary to \textit{create} inner bars, they can survive past that phase.  This appears to rule out the early models of \citet{friedli93}, where inner bars were transient phenomena triggered by gas inflows.

\section{Pattern Speeds}

The earliest theoretical arguments \citep{shlosman89,pfenniger90} and
models \citep[e.g.,][]{friedli93} suggested that inner bars would be
decoupled from --- and in fact faster-rotating than --- outer bars. 
Both orbital models (see the contribution by Witold Maciejewski in this
volume) and $N$-body simulations (see the contribution by Juntai Shen in
this volume) support this from a theoretical point of view.

%[Models of co-rotating: Shaw+93; Shlosman \& Heller]

From an observational point of view, there is good \textit{indirect}
evidence for decoupled inner bars.  The first observational studies of
double bars as a class \citep{bc93,friedli93} pointed out that inner and
outer bars seemed to be randomly oriented with respect to each other ---
in particular, inner bars were not found preferentially either
perpendicular or parallel to outer bars, which the simplest models of
corotating double bars would require.  There have been some more
sophisticated models of corotating inner bars \citep{shaw93,heller96},
but even these predict preferred orientations (the inner bar must lead
the outer bar).

\begin{figure*}[t!]
%\resizebox{\hsize}{!}{\includegraphics[clip=true]{pa-figure.eps}}
\resizebox{\hsize}{!}{\includegraphics[clip=true]{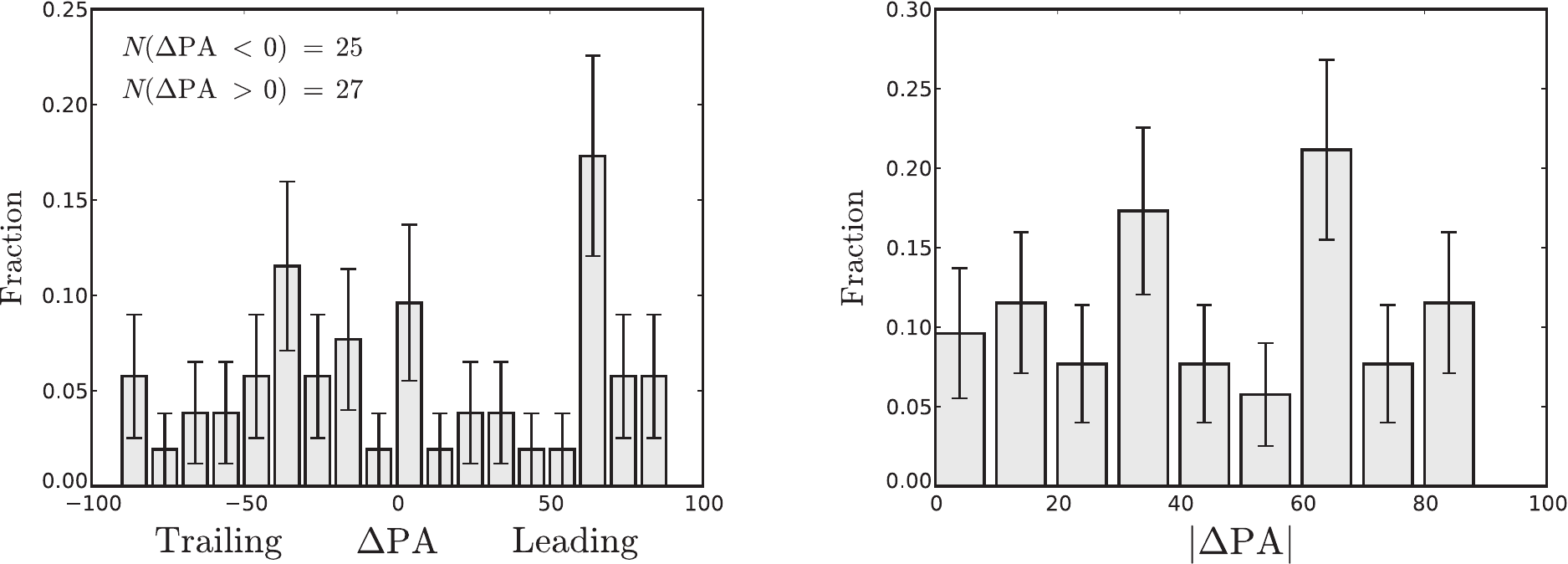}}
\caption{\footnotesize
Deprojected relative positions angles between the bars of
double-barred galaxies. Left: position angles in a leading/trailing
sense (positive = inner bar leads outer bar), for 52 galaxies where
sense of rotation can be determined. Right: absolute position angle
between inner and outer bars, for 61 galaxies.}
\label{fig:pa}
\end{figure*}

Figure~\ref{fig:pa} shows an updated plot of relative position angles
between inner and outer bars. As was seen earlier with smaller numbers,
inner bars do not preferentially lead or trail outer bars, and the
relative angles between them appear to be randomly distributed. This is
consistent with the general argument that inner bars rotate
independently. Although some recent models \citep{witold00,debattista07}
suggest that the relative patten speeds of inner bars should vary, so
that inner bars spend more time perpendicular to outer bars, this cannot
be a very strong effect, as inner bars are not preferentially seen in
near-perpendicular orientations.

%Demographics: Relative PAs from catalog [argument first made by bc93 [and friedli \& martinet?]]

% FIGURE: relative PAs between inner and outer bars
%    QUESTION: use both figures, or just |deltaPA| and mention
%    relative numbers of leading and trailing?
%

There is limited evidence for decoupled pattern speeds from
hydrodynamical modeling of individual galaxies, where one attempts to
match the gas flow in potentials with one or more rotating bars to the
observed gas kinematics of a particular galaxy. For example,
\citet{ann01} found a good match for gas morphology when the inner bar
of NGC~4314 was rotating faster than the outer bar; however, no
kinematic comparison was made.  More promising are the cases of NGC~1068
and NGC~4736, though complete, self-consistent modeling for both
galaxies is lacking \citep[see the summary in][]{erwin04}.

To date, there has been only one published attempt to \textit{directly}
measure pattern speeds in a double-barred galaxy, by \citet{corsini03}.
They applied the Tremaine-Weinberg (T-W) method to NGC 2950
(Figure~\ref{fig:n2950}), and were able to show that the two bars did
\textit{not} have the same pattern speed. While a pattern speed for the
outer bar was measured, determining a specific speed for the inner bar
proved more difficult.  The peculiar T-W results for the latter have
prompted arguments that the inner bar might actually be
\textit{counter}-rotating (\nocite{witold06}Maciejewski 2006; see also
\nocite{shen09}Shen \& Debattista 2009). While this is an intriguing
possibility, the existing models for forming counter-rotating inner bars
\citep{friedli96a,davies97} require the majority of stars in the inner
galaxy to counter-rotate, to the point of producing clear reversals in
the stellar rotation curve --- something not seen in NGC 2950 (compare
Figures 6 and 7 of Friedli 1996 with Figure~4 of Corsini et al.).  One
possible test of the counter-rotating inner-bar hypothesis would be to
compare hydrodynamical simulations  of prograde and retrograde inner-bar
systems with observed gas flows --- or even with observed dust lanes ---
since the resulting gas flows and shocks would presumably be rather
different.

\section{Double Bars and AGN?}

% No figures needed here

Because the original theoretical motivation for nested-bar systems was
to provide a mechanism for fuelling AGN \citep{shlosman89}, the question
of whether double bars actually enhance nuclear activity has been a
relatively popular one.

Most recent observational studies actually suggest that inner bars play
at best only a minor role in promoting nuclear activity.
\citet{martini99} and \citet{martini01} noted that nuclear bars were not
common in a small sample of local Seyfert galaxies; \citet{erwin02}
compared single- and double-barred galaxies and found no significant
difference in nuclear activity.

% martini99: nuclear bars not very common in Sy2, but nuclear spirals are

The study of \citet{laine02}, which used larger samples of Seyfert and
non-Seyfert galaxies, seems at first glance to indicate a correlation
between double bars and AGN: the fraction of galaxies in their Seyfert
sample with two bars is 21\%, versus only 13\% for the non-Seyfert
galaxies.  However, this is primarily a function of the higher overall
bar fraction in their Seyfert galaxies.  If we restrict ourselves to
\textit{barred} galaxies, then the fraction of (barred) Seyferts with
inner bars is not significantly higher than that for the control
galaxies (29\% vs.\ 25\%).

A comparison of bar frequencies for different bar sizes in the Laine et
al.\ study yields a curious result: galaxies hosting very small bars
(radii $< 1$ kpc), whether inner or not, are roughly as common in the
Seyfert and non-Seyfert samples (Kolmogorov-Smirnov $P = 0.67$). But
galaxies with very large bars (radii $> 3$ kpc) are far more likely to
be Seyferts (48\% of Sy galaxies, 21\% of control galaxies; K-S $P =
0.0051$). So \textit{if} bars fuel nuclei, it may only be \textit{large}
bars that matter.

\begin{acknowledgements}

I would like to thank the conference organizers for such an
excellent and stimulating workshop, as well as for the opportunity to
present this talk --- and, in addition, for the chance to see the
original site of the pioneering dissections by Vesalius and his
successors, which help inspire part of this talk. This work was
supported by Priority Programme 1177 of the Deutsche
Forschungsgemeinschaft.
%(``Witnesses of Cosmic History: Formation and
%Evolution of Black Holes, Galaxies and Their Environments'').

\end{acknowledgements}

\bibliographystyle{aa}

\end{document}